\documentclass[aps,pre,reprint,amsmath,amssymb,superscriptaddress,floatfix]{revtex4-2}

\usepackage{bm}
\usepackage{graphicx}
\usepackage{xcolor}
\usepackage[colorlinks=true,allcolors=blue]{hyperref}

\begin{document}

\title{Thresholdless IBW Emission and Alpha-to-Thermal-Ion Energy Channeling via Pole Resonance of Fusion-Product Ions }

\author{Hong Qin}
\affiliation{Department of Nuclear Engineering and Engineering Physics, University of Wisconsin-Madison, Madison, WI 53706, USA}
\affiliation{Department of Astrophysical Sciences, Princeton University, Princeton, NJ 08540, USA}
\author{Nathaniel J. Fisch}
\affiliation{Department of Astrophysical Sciences, Princeton University, Princeton, NJ 08540, USA}

\date{\today}

\begin{abstract}
Fusion reactions can release a substantial fraction, and in aneutronic reactions nearly all, of their energy as the kinetic energy of fusion-product ions. In magnetized plasmas, these energetic ions can form ring distributions in velocity space. Such distributions  can drive Dory-Guest-Harris (DGH)
electrostatic instabilities, but these self-instabilities require a finite
energetic-ion density and can be stabilized by a thermal background. We show
that the same background can instead enable a distinct instability mechanism: a
cyclotron pole of the minority energetic-ion susceptibility destabilizes a
stable ion Bernstein wave (IBW) eigenmode. When the energetic-ion harmonic is
distinct from the thermal-ion harmonics and its residue has the
destabilizing sign, exact root-pole resonance is
thresholdless in the ideal collisionless limit. A
species-resolved power balance shows that the energetic ions supply the free
energy while the thermal plasma receives it. In the LAPD proton--alpha
reference case at $24\ \mathrm{keV}$, $99.96\%$ of the alpha-particle power loss enters the coherent
proton response. This self-excited, ion-directed transfer provides a
possible linear building block for alpha-particle energy channeling in proton-Boron11 fusion.
\end{abstract}

\maketitle

\section{Introduction}
Energetic ion populations with ring-like perpendicular distributions arise naturally in magnetic mirrors~\cite{Post1966,Forest2024,Kolmes2024,Frank2025,Tran2025,Shih2026}, beam-injected plasmas, and plasmas containing fusion products. For $p$--${}^{11}$B fusion~\cite{Rostoker1993,Rostoker1997,Putvinski2019,Kolmes2022a,Ochs2022,wei2023,Kong2023,Liu2024,Ochs2024,Kamio2026} and other aneutronic fusion reactions using advanced fuels~\cite{Qin2024}, all fusion energy is carried by energetic ions. Separating alpha heating from isotopic mass effects and competing energy-transfer mechanisms has proved challenging in analyses of TFTR and early JET deuterium--tritium experiments~\cite{Budny2018AlphaHeating}. Efficient access to the energetic-ion free energy is especially important for alpha-particle channeling \cite{FischRax1992,Ochs2021,Ochs2022} and direct energy conversion \cite{Qin2025,Updike2025a}, motivating exploration of collective instabilities driven by energetic ions. Fusion-product-driven instabilities have been investigated as mechanisms for ion cyclotron emission~\cite{CottrellDendy1988,Dendy1994,Cook2013} and electron current drive~\cite{Cook2010,Cook2011}. The present study considers a cold-ring idealization, in which all energetic ions have the same perpendicular speed $v_{h0}$, so that the distribution function, normalized to $n_h$, can be written as
\begin{equation}
 f_{h0}^{\mathrm{ring}}(v_x,v_y,v_\parallel)
 =\frac{n_h}{\pi}\delta(v_x^2+v_y^2-v_{h0}^2)\delta(v_\parallel).
 \label{eq:ring-cartesian}
\end{equation}
The perpendicular distribution thus forms a ring in the $(v_x,v_y)$ plane. Such distributions can drive electrostatic cyclotron-harmonic instabilities of the type studied by Dory, Guest, and Harris (DGH)~\cite{Harris1959,Dory1965,Chen2026Control}.  In the ordinary DGH mechanism, the energetic-ion population  both establishes the collective mode and supplies its free energy. Consequently, the pure-ring instability studied by DGH requires a finite energetic-ion density. This threshold does not preclude a distinct instability driven by a dilute energetic-ion population.

Exciting the ordinary DGH instability becomes more difficult when the energetic ions form a minority population embedded in a dominant thermal plasma. The electron and thermal-ion responses screen the pure-ring collective response, and a DGH mode that is unstable in a sufficiently dense energetic-ion plasma can be stabilized as the thermal background is increased. One might therefore conclude that minority energetic ions, even if significantly above the ordinary DGH threshold, cannot generate coherent electrostatic emission.

In the present study, we show that the thermal background can in fact play the opposite role in driving an instability of a thermal-energetic system, but through a different mechanism. A thermal magnetized plasma supports stable ion Bernstein wave (IBW) eigenmodes~\cite{Bernstein1958,Stix1992}. A cyclotron pole of the minority energetic-ion susceptibility can destabilize one of these eigenmodes when the pole is tuned close to an IBW root, but remains distinct from the thermal-ion cyclotron poles. Although the mechanism is not designed as a free-space microwave source, it acts as a plasma analog of a resonant oscillator: minority energetic ions provide the free energy, while the thermal plasma supplies the wave cavity in the form of a stable IBW eigenmode. This is a root-pole resonance rather than a self-instability of the energetic-ion population. At exact noncommensurate resonance with a destabilizing pole residue, any nonzero minority density produces a complex-conjugate pair in the ideal collisionless problem, with a growth rate proportional to the square root of the minority density.

Moreover, a species-resolved power balance identifies the energetic ions as the energy source and
the thermal-ion component of the IBW as the dominant recipient. In the
proton--alpha case, the energetic alpha particles supply the power while nearly
all of their energy loss flows into the coherent thermal-proton response.

We first contrast the ordinary DGH instability with the stable thermal IBW
spectrum and with thermal stabilization of a DGH branch. We then derive
the mode-pole instability and test it with numerical calculations
based on Large Plasma Device (LAPD) parameters~\cite{Gekelman2016}. The
proton--alpha examples use a $2\%$ alpha particle ring near
$3\Omega_\alpha=1.5\Omega_p$. We examine the dependence on ring energy
both through wavenumber scans and at a fixed wavenumber. Finally,
we quantify the species-resolved power balance.

\section{General kinetic model}
For a homogeneous collisionless plasma in a uniform magnetic field $\bm B_0=B_0\hat{\bm z}$, we consider the electrostatic perturbation propagating strictly perpendicular to the magnetic field,
\begin{equation}
 \phi_1=\widetilde\phi\exp(ik_\perp x-i\omega t),
 \qquad k_\parallel=0.
 \label{eq:perturbation}
\end{equation}
The plasma consists of electrons, a dominant thermal-ion species $i$, and a
minority energetic-ion species $h$. For a single thermal-ion species and a single
energetic-ion species, quasineutrality gives
\begin{equation}
 n_e=Z_in_i+Z_hn_h.
 \label{eq:quasineutrality}
\end{equation}
The full dispersion relation is
\begin{equation}
 D(\omega,k_\perp)
 =1+\chi_e+\chi_i^{\mathrm{th}}+\chi_h^{\mathrm{ring}}=0.
 \label{eq:full-dispersion}
\end{equation}
Here, $\chi_s$  is the susceptibility of species $s$, and it measures how strongly it polarizes or screens the
electrostatic field. For  a gyrotropic equilibrium distribution $f_{s0}$, the strictly perpendicular
electrostatic susceptibility is
\begin{equation}
 \begin{aligned}
 \chi_s(\omega,k_\perp)
 ={}&\frac{q_s^2}{\epsilon_0m_sk_\perp^2}
 \sum_{n=-\infty}^{\infty}
 \frac{n\Omega_s}{\omega-n\Omega_s}\\
 &\times\int d^3v\,
 J_n^2\!\left(\frac{k_\perp v_\perp}{\Omega_s}\right)
 \frac{1}{v_\perp}
 \frac{\partial f_{s0}}{\partial v_\perp}.
 \end{aligned}
 \label{eq:generic-susceptibility}
\end{equation}
The velocity-space gradient of $f_{s0}$ determines the
sign of the response, the Bessel factor $J_n$ gives the finite-Larmor-radius dependent
coupling to harmonic $n$, and the poles at $\omega=n\Omega_s$ identify the
cyclotron resonances.

For each species $s$, we define
\begin{equation}
 \omega_{ps}^2=\frac{n_sq_s^2}{\epsilon_0m_s},
 \qquad
 \Omega_s=\frac{|q_s|B_0}{m_s}.
 \label{eq:frequencies}
\end{equation}
Here $n_s$, $q_s$, and $m_s$ are the density, charge, and mass,
respectively; $q_i=Z_i e$, $q_h=Z_h e$, and $q_e=-e$.

Both electrons and thermal ions have isotropic Maxwellian equilibria,
\begin{align}
 f_{s0}^{\mathrm{th}}(\bm v)
 &=\frac{n_s}{(2\pi v_{ts}^2)^{3/2}}
 \exp\left[-\frac{v_\perp^2+v_\parallel^2}{2v_{ts}^2}\right],
 \nonumber\\
 v_{ts}^2&=\frac{T_s}{m_s},\qquad s=e,i.
 \label{eq:maxwellian}
\end{align}
With
\begin{equation}
 \rho_s=\frac{v_{ts}}{\Omega_s},
 \qquad b_s=k_\perp^2\rho_s^2,
 \qquad \Gamma_\ell(b_s)=e^{-b_s}I_\ell(b_s),
 \label{eq:thermal-definitions}
\end{equation}
the strictly perpendicular Maxwellian susceptibility is
\begin{equation}
 \chi_s^{\mathrm{th}}(\omega,k_\perp)
 =-\frac{2\omega_{ps}^2}{b_s}
 \sum_{\ell=1}^{\infty}
 \frac{\ell^2\Gamma_\ell(b_s)}{\omega^2-\ell^2\Omega_s^2},
 \qquad s=e,i.
 \label{eq:thermal-susceptibility}
\end{equation}
Here, $I_\ell$ is a modified Bessel function. Equation~\eqref{eq:thermal-susceptibility} makes explicit that the
perpendicular susceptibility depends on frequency only through $\omega^2$.
The same dependence follows from Eq.~\eqref{eq:generic-susceptibility} by
pairing the $n$ and $-n$ terms. We therefore introduce
$z\equiv\omega^2$ here and henceforth formulate the dispersion relation in
the complex $z$ plane.

In the small-electron-gyroradius
limit, $\Gamma_1(b_e)=b_e/2+O(b_e^2)$ while
$\Gamma_\ell(b_e)=O(b_e^\ell)$ for $\ell\ge2$, and
\begin{equation}
 \chi_e^{\mathrm{th}}
 =-\frac{\omega_{pe}^2}{z-\Omega_e^2}+O(b_e)
 \simeq\frac{\omega_{pe}^2}{\Omega_e^2}+O(b_e),
 \label{eq:electron-cold-limit}
\end{equation}
where the second form also uses $|\omega|\ll\Omega_e$. The thermal
electron response reduces continuously to the cold magnetized response
when $b_e\ll1$. Equation~\eqref{eq:thermal-susceptibility} also shows
that all Maxwellian cyclotron poles have negative residues in the $z$ plane.

For the ring distribution of an energetic species, its susceptibility is~\cite{Dory1965,Qin2007}
\begin{align}
 \chi_h^{\mathrm{ring}}(z,k_\perp)
 &=\sum_{n=1}^{\infty}\frac{\mathcal R_n(k_\perp)}
 {z-n^2\Omega_h^2},\nonumber\\
 \mathcal R_n
 &=-\frac{2\omega_{ph}^2}{\lambda_h}n^2
 \frac{dJ_n^2(\lambda_h)}{d\lambda_h}.
 \label{eq:ring-residue-model}
\end{align}
where $\mathcal R_n$ is the actual residue of the pole in the $z$ plane, and
\begin{equation}
 \lambda_h=\frac{k_\perp v_{h0}}{\Omega_h},
 \label{eq:ring-parameter}
\end{equation}
Unlike the positive thermal-ion quantities introduced below, it is sign
indefinite because $dJ_n^2/d\lambda_h=2J_nJ_n'$ changes sign with
$\lambda_h$. The susceptibility expressions can also be obtained from the
general energetic magnetized susceptibility tensor~\cite{Qin2007}.

Unless otherwise specified, the numerical examples use the same LAPD-based reference parameters:
$B_0=0.05\ \mathrm{T}$, a $24\ \mathrm{keV}$ perpendicular alpha ring,
a cold perpendicular low-frequency electron response with nominal
$T_e=10\ \mathrm{eV}$, and $5\ \mathrm{eV}$ background ions when
present. Frequencies are normalized to
$\Omega_p=eB_0/m_p$, even in the cases without physical protons,
and $n_{\alpha0}=2.0\times10^{15}\ \mathrm{m}^{-3}$ denotes the
reference alpha density. In these numerical calculations the ion masses
are approximated by $m_p=u$, $m_D=2u$, and $m_\alpha=4u$, where $u$ is
the atomic mass unit. The cases differ in background-ion species and
density, alpha density and ring energy, and $k_\perp$, as specified in
the text.

\section{Ordinary DGH self-instability}
We first remove the thermal ions and consider the pure-ring dispersion relation
\begin{equation}
 D_{\mathrm{DGH}}(z,k_\perp)
 =1+\chi_e+\chi_h^{\mathrm{ring}}=0.
 \label{eq:dgh-dispersion}
\end{equation}
In this system, all ion-frequency structure is supplied by $\chi_h^{\mathrm{ring}}$. The resulting unstable modes are therefore self-modes of the energetic-ion population.

In the limit $n_h\to0$, both the energetic-ion and neutralizing-electron
susceptibilities vanish in proportion to $n_h$, and each root remaining
near a ring pole is a real perturbation of that pole. Complex roots can
therefore appear only after two real roots coalesce at a finite density
$n_h^{\mathrm{DGH}}$~\cite{Harris1959,Dory1965}. This bifurcation is the
spectral signature of spontaneous parity-time ($\mathcal{PT}$) symmetry
breaking in the linearized collisionless system. In the unbroken phase,
$\mathcal{PT}$ symmetry constrains the eigenfrequencies to be real; at an
exceptional point, two real eigenmodes coalesce, and after symmetry breaking
they emerge as a complex-conjugate pair~\cite{Bender1998,Bender2024,
Qin2019KH,Fu2020KH,Zhang2020PT,Qin2021,Qin2026PTWaves}.

\begin{figure}[t]
 \centering
 \includegraphics[width=\columnwidth]{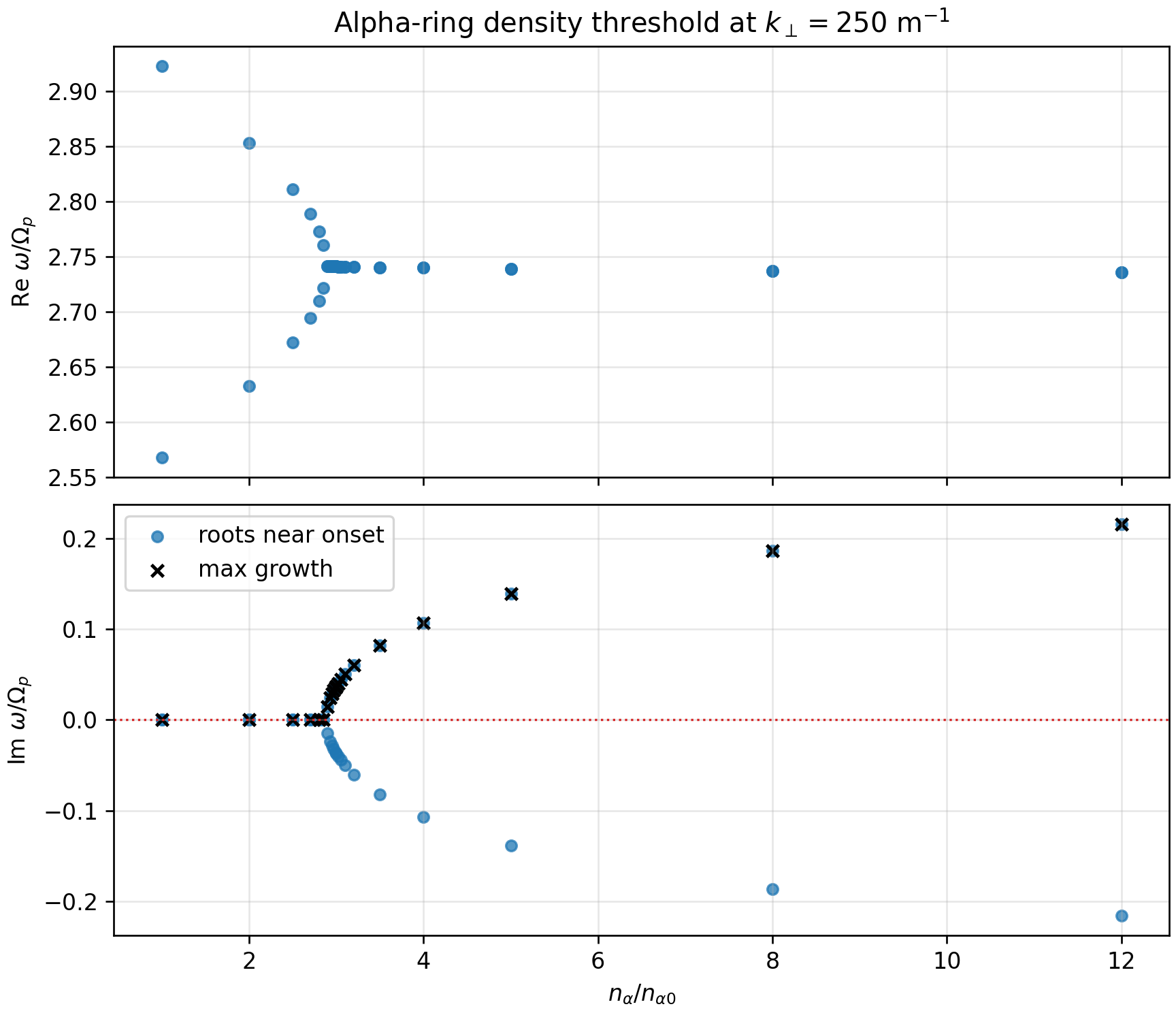}
 \caption{Ordinary alpha-ring DGH self-instability without thermal ions.
 Top: real parts of roots near onset. Bottom: imaginary parts; black
 crosses mark the maximum growth rate at each density. The instability
 threshold is $n_\alpha/n_{\alpha0}=2.90$.}
 \label{fig:dgh}
\end{figure}

Figure~\ref{fig:dgh} demonstrates this finite threshold for the common
LAPD-based alpha ring with no thermal ions. At fixed $k_\perp=250\ \mathrm{m}^{-1}$, the alpha
density is scanned relative to $n_{\alpha0}$. The ordinary DGH
instability appears at $n_\alpha/n_{\alpha0}=2.90$.

\section{Stable thermal ion Bernstein waves}
We next remove the energetic ions and consider the standard IBW in a plasma with thermal ions and electrons.  The thermal dispersion function is
\begin{equation}
 D_{\mathrm{th}}(z,k_\perp)
 =1-\sum_{\ell=1}^{\infty}
 \frac{E_\ell(k_\perp)}{z-\ell^2\Omega_e^2}
 -\sum_{m=1}^{\infty}\frac{A_m(k_\perp)}{z-m^2\Omega_i^2},
 \label{eq:thermal-dispersion}
\end{equation}
where
\begin{align}
 E_\ell(k_\perp)
 &=\frac{2\omega_{pe}^2}{b_e}\ell^2e^{-b_e}I_\ell(b_e)>0,
 \nonumber\\
 A_m(k_\perp)
 &=\frac{2\omega_{pi}^2}{b_i}m^2e^{-b_i}I_m(b_i)>0.
 \label{eq:thermal-residue}
\end{align}
The positivity of $E_\ell$ and $A_m$ gives a direct stability proof. For $z=x+iy$,
\begin{align}
 \operatorname{Im}D_{\mathrm{th}}(z,k_\perp)
 ={}&y\sum_{\ell=1}^{\infty}
 \frac{E_\ell}{(x-\ell^2\Omega_e^2)^2+y^2}
 \nonumber\\
 &+y\sum_{m=1}^{\infty}
 \frac{A_m}{(x-m^2\Omega_i^2)^2+y^2}.
 \label{eq:thermal-imaginary}
\end{align}
Every coefficient of $y$ is positive. Hence $D_{\mathrm{th}}=0$ requires $y=0$. Moreover, $D_{\mathrm{th}}(z)>1$ for $z\leq0$, so every zero has $z>0$ and therefore real $\omega$ in the ideal perpendicular collisionless problem. For real $z$ away from the poles,
\begin{equation}
 D_{\mathrm{th}}'(z,k_\perp)
 =\sum_{\ell=1}^{\infty}
 \frac{E_\ell}{(z-\ell^2\Omega_e^2)^2}
 +\sum_{m=1}^{\infty}
 \frac{A_m}{(z-m^2\Omega_i^2)^2}>0.
 \label{eq:thermal-slope}
\end{equation}
Thus, every thermal IBW root away from a cyclotron pole is simple and has positive slope. We denote a selected root by
\begin{equation}
 D_{\mathrm{th}}(z_t,k_\perp)=0.
 \label{eq:thermal-root}
\end{equation}

\section{Thermal stabilization of minority DGH modes}
We now add thermal deuterium to an alpha ring that is already DGH
unstable. Since $\Omega_D=\Omega_\alpha=\Omega_p/2$, their cyclotron
harmonics are commensurate. The study in this section tests the thermal stabilization of the ordinary DGH branch, not the noncommensurate
root-pole mechanism developed below.

\begin{figure}[t]
 \centering
 \includegraphics[width=\columnwidth]{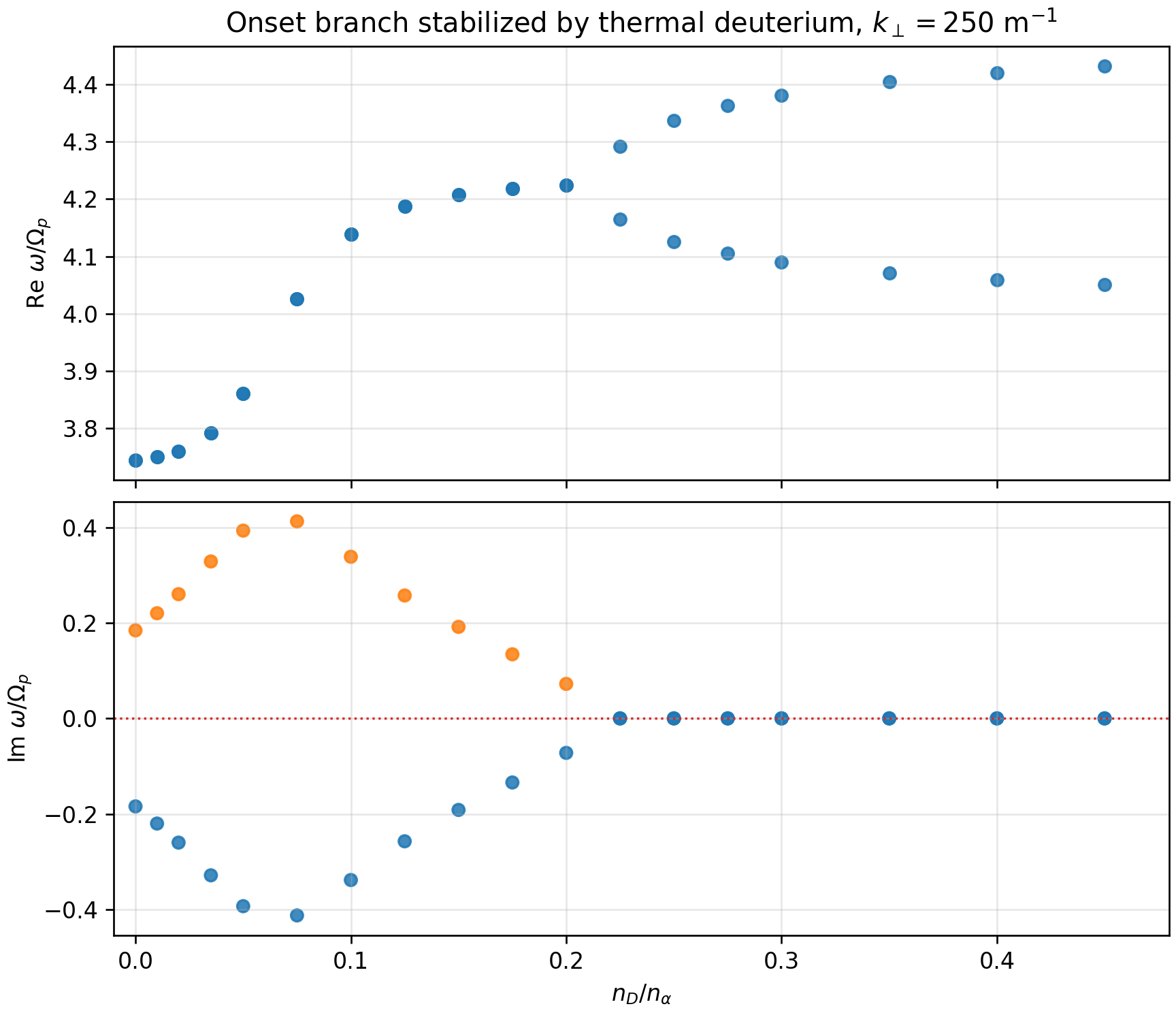}
 \caption{Stabilization of the ordinary alpha-ring DGH instability by
 thermal deuterium. Top: real parts of the tracked roots. Bottom:
 imaginary parts, with orange and blue denoting the growing and damped
 members before stabilization. The stabilization threshold is
 $n_D/n_\alpha=0.23$.}
 \label{fig:thermal-stabilization}
\end{figure}

At fixed $k_\perp=250\ \mathrm{m}^{-1}$ and
$n_\alpha=5n_{\alpha0}$, we increase $n_D$ while enforcing
$n_e=n_D+2n_\alpha$. As shown in Fig.~\ref{fig:thermal-stabilization}, the tracked branch grows nonmonotonically at low
$n_D/n_\alpha$, then its conjugate imaginary parts collapse to zero. The
thermal-deuterium stabilization threshold is $n_D/n_\alpha=0.23$;
above it the pair has split into two real roots.

The stabilization can also be understood perturbatively in the minority limit. Away from an energetic-ion pole, $\chi_h^{\mathrm{ring}}$ is a regular real function of real $z$. A sufficiently small $n_h$ therefore gives only a real perturbation of the simple thermal roots established by Eq.~\eqref{eq:thermal-slope}. A complex pair cannot emerge from an isolated simple root under such a regular infinitesimal perturbation.

This regular-perturbation argument ceases to be valid when an energetic-ion
cyclotron pole approaches a thermal IBW root. Even though the pole
residue vanishes with $n_h$, its denominator vanishes at the same point,
so the minority susceptibility cannot be treated as a small smooth
correction. The thermal root and the nearby singularity must instead be
retained together in a local dispersion relation. Their interaction
produces a root-pole splitting that can be complex for arbitrarily small
nonzero residue at exact resonance, leading to the mechanism developed in
the next section.

\section{New instability mechanism: mode-pole instability driven by minority energetic ions}
The $n$th energetic-ion cyclotron pole is located at
\begin{equation}
 z_{hn}=n^2\Omega_h^2.
 \label{eq:energetic-pole}
\end{equation}
The new mechanism requires a thermal IBW root close to this pole,
\begin{equation}
 z_t(k_\perp)\simeq z_{hn},
 \label{eq:resonance}
\end{equation}
while the energetic-ion harmonic remains distinct from the thermal-ion cyclotron harmonics,
\begin{equation}
 n\Omega_h\ne m\Omega_i.
 \label{eq:noncommensurate}
\end{equation}
Condition~\eqref{eq:noncommensurate} is essential. If the two harmonics coincide, the energetic ions merely modify the residue of an existing thermal-ion pole rather than provide a distinct pole near an IBW root.

To expose the local structure, we start from the expression of $ \chi_h^{\mathrm{ring}}(z,k_\perp)$ given by Eq.~\eqref{eq:ring-residue-model} and  separate the selected pole and absorb all nonresonant energetic-ion harmonics into a regular dispersion function,
\begin{equation}
 D(z,k_\perp)
 =D_{\mathrm{reg}}(z,k_\perp)
 +\frac{\mathcal R_n(k_\perp)}{z-z_{hn}}.
 \label{eq:regular-separation}
\end{equation}
Let $z_r$ denote the weakly shifted regular IBW root,
\begin{equation}
 D_{\mathrm{reg}}(z_r,k_\perp)=0,
 \qquad z_r=z_t+O(n_h).
 \label{eq:regular-root}
\end{equation}
Near this root, Eq.~\eqref{eq:regular-separation} reduces to
\begin{equation}
 D_{\mathrm{reg}}'(z_r)(z-z_r)(z-z_{hn})+\mathcal R_n=0.
 \label{eq:local-dispersion}
\end{equation}
The two local roots are
\begin{equation}
 z_\pm=\frac{z_r+z_{hn}}{2}
 \pm\frac{1}{2}
 \sqrt{(z_r-z_{hn})^2-
 \frac{4\mathcal R_n}{D_{\mathrm{reg}}'(z_r)}}.
 \label{eq:local-roots}
\end{equation}
For a dilute energetic-ion population, $D_{\mathrm{reg}}'(z_r)>0$ by continuity from Eq.~\eqref{eq:thermal-slope}. The destabilizing residue condition is therefore
\begin{equation}
 \mathcal R_n>0.
 \label{eq:residue-condition}
\end{equation}
The roots form a complex-conjugate pair when
\begin{equation}
 (z_r-z_{hn})^2
 <\frac{4\mathcal R_n}{D_{\mathrm{reg}}'(z_r)}.
 \label{eq:instability-condition}
\end{equation}

At exact resonance, $z_r=z_{hn}$, Eq.~\eqref{eq:local-roots} becomes
\begin{equation}
 z_\pm=z_{hn}\pm i
 \sqrt{\frac{\mathcal R_n}{D_{\mathrm{reg}}'(z_r)}}.
 \label{eq:exact-resonance-roots}
\end{equation}
Because $\mathcal R_n\propto\omega_{ph}^2\propto n_h$, the weak-growth limit gives
\begin{equation}
 \gamma\simeq\frac{1}{2\sqrt{z_{hn}}}
 \sqrt{\frac{\mathcal R_n}{D_{\mathrm{reg}}'(z_r)}}
 \propto\sqrt{n_h}.
 \label{eq:growth-scaling}
\end{equation}
Thus, for a destabilizing residue, the growth rate is positive at every
nonzero minority density and tends to zero as $n_h\to0$; there is no
density threshold at exact resonance. At finite detuning
$\Delta z=z_r-z_{hn}$, Eq.~\eqref{eq:instability-condition} instead gives
$n_{h,\mathrm{crit}}\propto(\Delta z)^2$ in the local approximation.

\section{Mode-pole instability of thermal proton-energetic alpha system}
We now specialize the common numerical setup to a thermal proton
background with
\begin{equation}
 n_p=1.0\times10^{17}\ \mathrm{m}^{-3},
 \qquad
 \frac{n_\alpha}{n_p}=0.02.
 \label{eq:lapd-parameters}
\end{equation}
The third alpha harmonic satisfies
\begin{equation}
 3\Omega_\alpha=1.5\Omega_p,
 \label{eq:alpha-harmonic}
\end{equation}
and is therefore distinct from every proton cyclotron harmonic. The thermal
proton IBW dispersion contains a real root at $\omega=1.5\Omega_p$ for a
resonant perpendicular wavenumber $k_*$ determined from
$D_{\mathrm{th}}(1.5^2\Omega_p^2,k_*)=0$.  For the minority-free thermal
background, $k_*\simeq251.06\ \mathrm{m}^{-1}$.

\begin{figure*}[t]
 \centering
 \includegraphics[width=\textwidth]{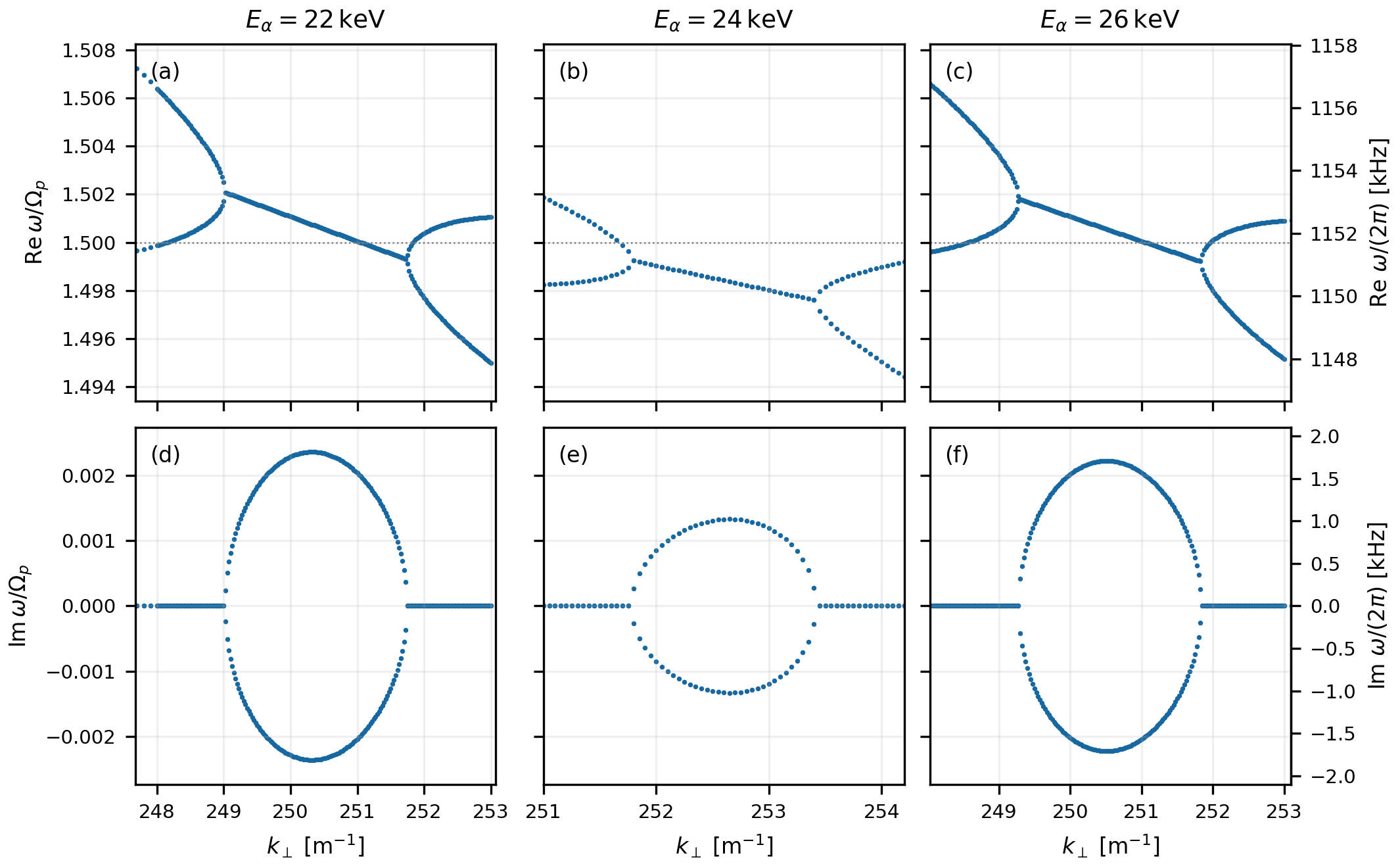}
 \caption{Proton--alpha roots of the full perpendicular dispersion
 relation near the third alpha harmonic for $E_\alpha=22$, $24$, and
 $26\ \mathrm{keV}$, with $n_\alpha/n_p=0.02$ and the
 background parameters fixed. Panels (a)--(c) show real frequency and
 panels (d)--(f) show imaginary frequency, normalized to $\Omega_p$;
 the right-hand axes give the corresponding values of $\omega/(2\pi)$
 in kilohertz. Positive imaginary frequency denotes instability.
 The horizontal ranges differ among energies and are chosen so that the
 unstable interval occupies approximately half of each panel.}
 \label{fig:lapd-spectrum}
\end{figure*}

Figure~\ref{fig:lapd-spectrum} shows wavenumber scans of the full
dispersion relation through this noncommensurate crossing for alpha-ring
energies of $22$, $24$, and $26\ \mathrm{keV}$, with the remaining
parameters identical. 

The wavenumber scans  show that the
complex-conjugate pair occupies a finite interval whose position and
width depend on the alpha-ring energy. Outside each unstable interval
the two roots are real, as predicted by
Eq.~\eqref{eq:instability-condition}. The growth rates peak at
$k_\perp=250.3$, $252.7$, and $250.5\ \mathrm{m}^{-1}$ for  $E_\alpha=22$, $24$, and
 $26\ \mathrm{keV}$, respectively.

\begin{figure}[t]
 \centering
 \includegraphics[width=\columnwidth]{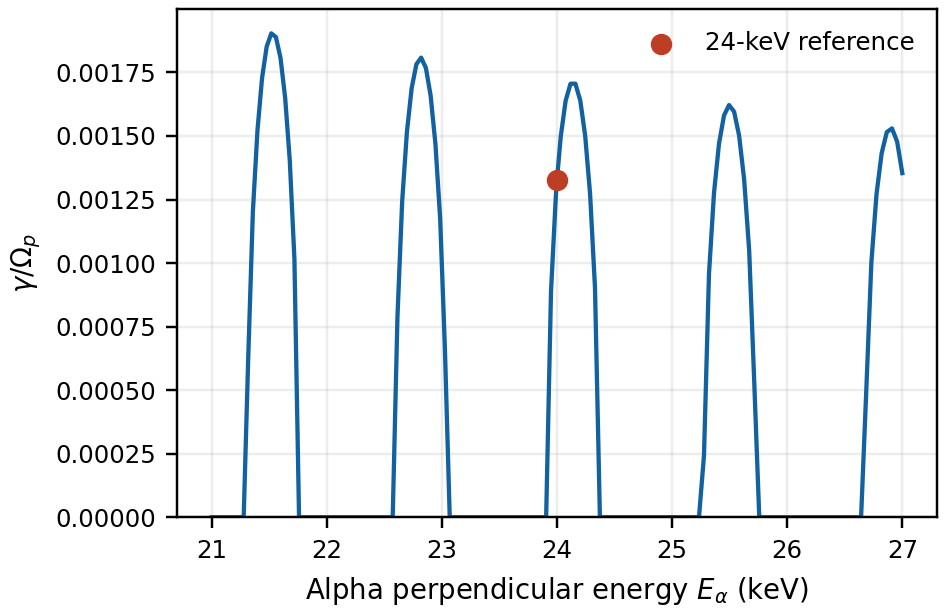}
 \caption{Growth rate of the local proton--alpha root pair near
 $3\Omega_\alpha$ as the alpha-ring energy varies from $21$ to
 $27\ \mathrm{keV}$. The perpendicular wavenumber is fixed at
 $252.7\ \mathrm{m}^{-1}$ and $n_\alpha/n_p=0.02$; all other background
 parameters are unchanged. The marked point is the reference case.}
 \label{fig:alpha-energy}
\end{figure}

\begin{figure}[t]
 \centering
 \includegraphics[width=\columnwidth]{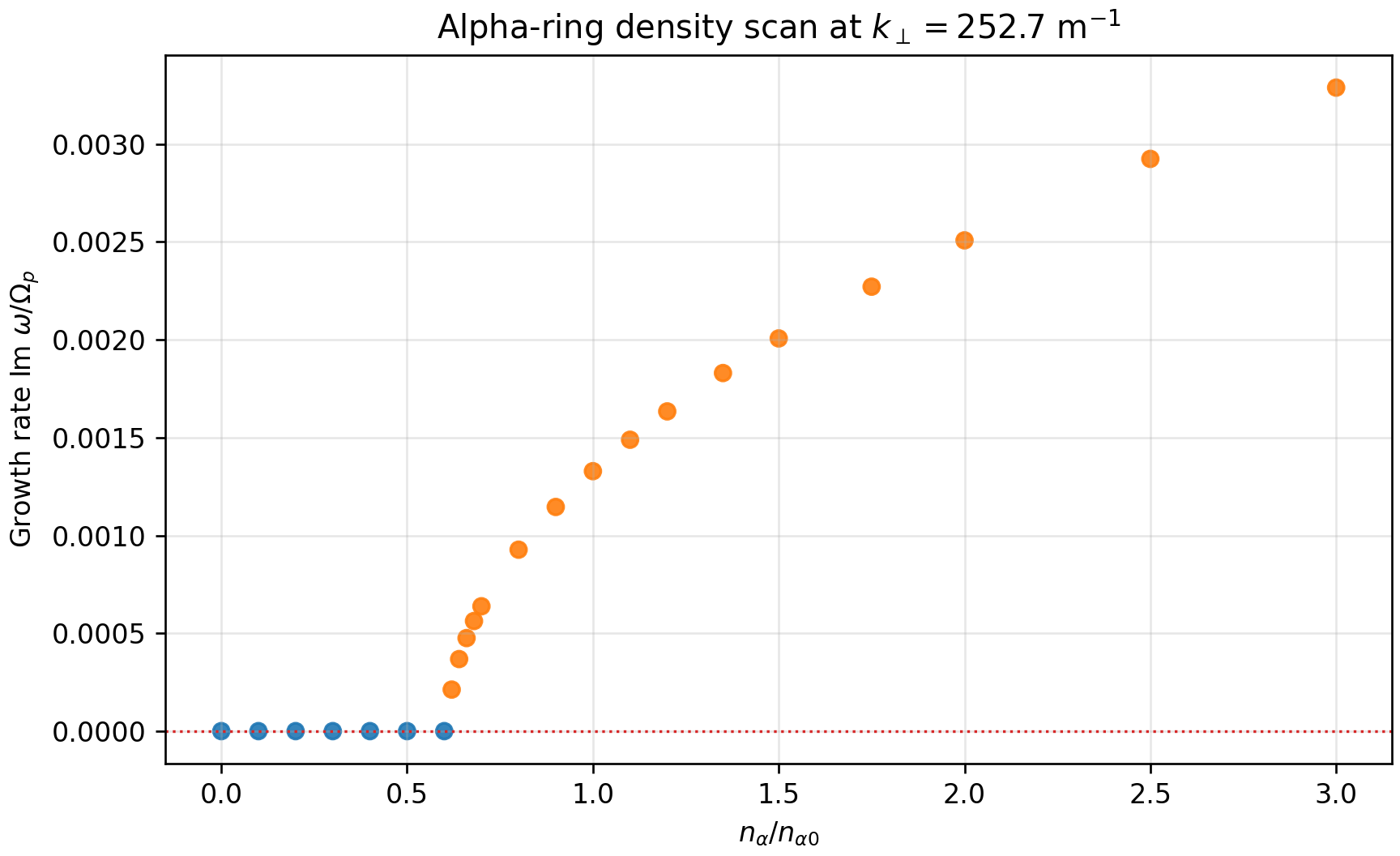}
 \caption{Growth rate of the fixed-wavenumber proton--alpha mode versus
 alpha density. Blue and orange points denote stable and unstable sampled
 roots, respectively. The finite detuned onset is
 $n_\alpha/n_{\alpha0}=0.62$; this is not the exact-resonance
 thresholdless limit.}
 \label{fig:alpha-density}
\end{figure}

Figure~\ref{fig:alpha-energy} shows the growth rate as a function of  $E_\alpha$ at $k_\perp=252.7\ \mathrm{m}^{-1}$, which maximizes the growth rate at   $E_\alpha=24 \ \mathrm{keV}$.  The dependence is nonmonotonic, with alternating
stable and unstable intervals.  In particular, the $22$- and
$26\ \mathrm{keV}$ cases are stable for this pair at the fixed
$k_\perp=252.7\ \mathrm{m}^{-1}$, but become unstable at other
wavenumbers, as shown in Fig.~\ref{fig:lapd-spectrum}.
Figure~\ref{fig:alpha-density} fixes the alpha-ring energy at
$24\ \mathrm{keV}$ and
$k_\perp=252.7\ \mathrm{m}^{-1}$ and varies only the alpha density with
the proton background unchanged. The detuned mode becomes unstable at
$n_\alpha/n_{\alpha0}=0.62$.

At this fixed wavenumber the regular thermal root and third-alpha pole are
slightly detuned, so a finite density is required to satisfy
Eq.~\eqref{eq:instability-condition}. This behavior is consistent with
the local prediction
$n_{\alpha,\mathrm{crit}}\propto(\Delta z)^2$, although a scan over
multiple detunings is required to verify the quadratic boundary. The
thresholdless limit is recovered only when the system is retuned to exact
resonance, $\Delta z=0$, with the destabilizing residue sign
$\mathcal R_3>0$.

These scans identify the unstable branch as a thermal proton IBW energized
by the third-alpha pole. We next quantify the species-resolved power flow.

\section{Energy transfer to thermal ions}
We use peak phasors,
$\bm E_{\rm phys}=\operatorname{Re}[\bm E\exp(i\bm k\cdot\bm x-i\omega t)]$,
with $\omega=\omega_r+i\gamma$, so that the vacuum electrostatic energy
density is $W_E=\epsilon_0|E|^2/4$.  The time-averaged power delivered by
the field to species $s$ is
\begin{equation}
 P_s=\frac{1}{2}\operatorname{Re}(\bm J_s\cdot\bm E^*)
 =\frac{\epsilon_0|E|^2}{2}\operatorname{Re}[-i\omega\chi_s].
 \label{eq:species-power}
\end{equation}
For a thermal species, each cyclotron harmonic has the form
$\chi_{sm}^{\rm th}=-A_{sm}/(\omega^2-z_{sm})$, with $A_{sm}>0$ and
$z_{sm}=m^2\Omega_s^2$.  Equation~\eqref{eq:species-power} then gives
\begin{equation}
 P_{sm}^{\rm th}=\frac{\epsilon_0|E|^2}{2}\gamma A_{sm}
 \frac{|\omega|^2+z_{sm}}{|\omega^2-z_{sm}|^2}>0.
 \label{eq:thermal-power}
\end{equation}
Both the thermal-ion and thermal-electron responses absorb power from
a growing mode.  In contrast, the selected energetic pole gives
\begin{equation}
 P_{h,n}=-\frac{\epsilon_0|E|^2}{2}\gamma\mathcal R_n
 \frac{|\omega|^2+z_{hn}}{|\omega^2-z_{hn}|^2},
 \label{eq:energetic-power}
\end{equation}
thus the destabilizing condition $\mathcal R_n>0$ is precisely the condition
$P_{h,n}<0$: the energetic population does work on the wave.  The dispersion
relation gives the exact energy balance of the growing eigenmode,
\begin{equation}
 P_e+P_i+P_h=-2\gamma W_E,
 \label{eq:exact-power-balance}
\end{equation}
which shows that the loss of energetic particle energy $L_h=-P_h>0$ is
converted to field energy and energy in the coherent responses of the
thermal ions and electrons.

Near exact resonance, let $z_0=z_{hn}$ and define the dimensionless thermal
weights
\begin{equation}
 H_s(z_0)=\sum_m A_{sm}\frac{z_0+z_{sm}}{(z_0-z_{sm})^2}>0.
 \label{eq:thermal-weight}
\end{equation}
Since $\gamma=O(\sqrt{n_h})$, the leading thermal powers are
$P_s=2\gamma W_EH_s+O(n_h^{3/2})$.  Equation \eqref{eq:exact-power-balance} yields
\begin{equation}
 \begin{aligned}
 \frac{2\gamma W_E}{L_h}&=\frac{1}{1+H_e+H_i},\\
 \frac{P_e}{L_h}&=\frac{H_e}{1+H_e+H_i},&
 \frac{P_i}{L_h}&=\frac{H_i}{1+H_e+H_i}.
 \end{aligned}
 \label{eq:partition-fractions}
\end{equation}
The energy partition is therefore a property of the resonant thermal
eigenmode, not of the small minority density.  Ion-directed transfer occurs
when $H_i\gg1+H_e$.  For a proton IBW, $H_e\simeq\omega_{pe}^2/\Omega_e^2$ whereas
$H_p\sim(\omega_{pp}^2/\Omega_p^2)F_p$, where $F_p$ is the finite-Larmor-radius
harmonic factor.  More precisely, with $x_0=z_0/\Omega_p^2$,
\begin{equation}
 F_p(b_p,x_0)=\frac{2}{b_p}\sum_{m=1}^{\infty}m^2\Gamma_m(b_p)
 \frac{x_0+m^2}{(x_0-m^2)^2},
 \label{eq:proton-flr-factor}
\end{equation}
so that $H_p=(\omega_{pp}^2/\Omega_p^2)F_p$ exactly for the Maxwellian
susceptibility in Eq.~\eqref{eq:thermal-susceptibility}. Consequently
$H_p/H_e\sim(m_p/m_e)F_p\gg1$ except where $F_p$ is anomalously small.

\begin{samepage}
For the $24\ \mathrm{keV}$ LAPD reference case in
Eq.~\eqref{eq:lapd-parameters}, we evaluate the power at the same
$k_\perp=252.7\ {\rm m}^{-1}$ used in the density and energy scans,
corresponding to $b_p\simeq1.324$. The unstable root is
$\omega/\Omega_p=1.498325+1.328365\times10^{-3}i$.
In units of $P_0=2\gamma W_E$, the species powers are
\begin{equation}
 \frac{P_\alpha}{P_0}=-11761.014,\
 \frac{P_e}{P_0}=4.280,\
 \frac{P_p}{P_0}=11755.734.
 \
 \label{eq:lapd-powers}
\end{equation}
\end{samepage}
These values satisfy $P_\alpha+P_e+P_p+P_0=0$. Normalized instead to $L_\alpha=-P_\alpha$, the field and electron shares are
$8.50\times10^{-5}$ and $3.64\times10^{-4}$, respectively, while
$99.96\%$ enters the coherent proton component of the IBW.  These numbers
characterize this finite-density reference mode; they are not asserted to
hold at every energy or wavenumber in the scans. 
Moreover, coherent proton-wave energy is not yet irreversible proton heat.
Collisions, phase mixing, trapping, damping, or nonlinear mode conversion
must thermalize it, but the linear instability determines that the available
power is overwhelmingly deposited in the thermal-ion part of the wave.

\section{Discussion and conclusion}
The mode-pole instability is distinct from the ordinary DGH
self-instability \cite{Harris1959,Dory1965}. A pure ring distribution in the velocity space becomes unstable only above a finite density,
and a thermal-ion background can suppress that branch. The same thermal
plasma, however, supplies stable IBW eigenmodes. When a simple IBW root is
tuned to a distinct positive-residue energetic-ion pole, the pair becomes complex.
Exact noncommensurate resonance is thresholdless in the ideal collisionless
limit and gives $\gamma\propto\sqrt{n_h}$, whereas finite detuning
produces the finite onset observed in Fig.~\ref{fig:alpha-density},
consistent with $n_{h,\mathrm{crit}}\propto(\Delta z)^2$.

For the LAPD thermal proton--energetic alpha examples, a $2\%$ alpha ring destabilizes the
thermal proton IBW near $3\Omega_\alpha=1.5\Omega_p$ over energy-dependent
wavenumber intervals. For the $24\ \mathrm{keV}$
reference mode, the alpha particles
supply the power, while $99.96\%$ of their loss enters the coherent
proton response; the thermal plasma is therefore both the wave cavity and
the dominant recipient. Coherent proton-wave energy is not yet irreversible
heat, which requires collisions, phase mixing, trapping, damping, or
nonlinear conversion.

This self-excited transfer of energy from energetic ions to another ion species is a form of alpha channeling, an effect originally intended for tokamaks~\cite{FischRax1992}, but which can also occur in a mirror geometry \cite{fisch06} such as LAPD.  In the case of self-excitation, once the instability overcomes losses, the alpha particle population supplies the wave power without a large external RF drive.
Controlled phase-space diffusion and selective damping are still required. Alpha-channeling via stimulated emission of fast Alfv\'en waves has also been demonstrated in kinetic simulations~\cite{Cook2017}.
In particular, since all the fusion energy released in a $p$--${}^{11}$B fusion reaction is carried by alpha particles, channeling their energy to protons would be particularly useful in hybrid fast--thermal concepts \cite{Kolmes2022a,Ochs2022,Ochs2024}.
However, a simple one-dimensional plateau-relaxation estimate illustrates a limitation. For a monoenergetic perpendicular ring $v_x^2+v_y^2=v_{h0}^2$, the reduced distribution in $v_x$ is $f(v_x)\propto(v_{h0}^2-v_x^2)^{-1/2}$. Complete flattening over $-v_{h0}<v_x<v_{h0}^2$ changes $\langle v_x^2\rangle$ from $v_{h0}^2/2$ to $v_{h0}^2/3$ and therefore extracts only $v_{h0}^2/6$, or one sixth of the initial perpendicular kinetic energy.  This limits the utility of the energy transfer, at least if performed directly with one wave.  There is the possibility, however, that the waves identified here that can release energy might be useful if used in conjunction with other waves \cite{fisch1995alpha2}.

Because of the quasilinear saturation of the direct energy transfer, an application more immediately useful than the fusion reactivity produced by this energy transfer between ion populations might be in producing current drive effects \cite{fisch87}.
To do so requires producing an asymmetry in the direction parallel to the magnetic field, such as by injecting waves traveling in one direction with respect to the magnetic field. Since the energy transfer requires a sharp frequency resonance, it may be imagined that ions moving in one parallel direction may experience the resonance due to a Doppler shift, whereas ions moving in the counter-streaming direction do not.
In that case, there could be perpendicular energy transfer from high-energy alpha particles to a lower-energy species.
If the other ion species has a different ion charge state, then conditions for an ion current drive effect are set up \cite{fisch81b}.

To see this, suppose an amount of perpendicular energy is resonantly channeled from minority high-energy alpha particles to lower-energy hydrogen, such that all resonant ions are moving in one parallel direction.
Consider ion-ion collisions only.
A resonant energetic alpha particle moving in one direction is slowed down by bulk ions.
If it is slowed down in the perpendicular energy, it collides more with the bulk ions, losing its parallel momentum faster to the bulk ions.
This creates a net drift of bulk ions in the direction of the resonant alpha particle and, by momentum conservation, alpha particles must move on average in the opposite direction.
Consider then the frame of reference in which the total ion current is zero.
If the bulk ion species has a different ion charge state than the alpha particles, then electrons will be dragged in the direction of the higher-Z drifting species, creating net current in the frame of reference in which the ion current vanishes, or any frame of reference, since current is frame invariant for a quasi-neutral plasma.
This is exactly the {\it inverse} of the minority ion species current drive effect, where minority ions are heated rather than cooled \cite{fisch81b}.
The opposite effect would occur with alpha particle damping of perpendicularly propagating waves, an effect thought to limit lower hybrid current drive \cite{fisch92b}.

An important limitation of the present analysis is the assumption of
strictly perpendicular propagation, $k_\parallel=0$. In a finite plasma,
parallel electron motion can modify the ion Bernstein wave spectrum and
the resonance with the energetic-ion cyclotron harmonic, even when the
propagation is nearly perpendicular. The growth rates might be sensitive to changes in this resonance. Quantifying this dependence requires a finite-$k_\parallel$ kinetic analysis, which is left for future work. 

Regardless of applications, the present work reveals a new energetic-ion-driven instability distinct from the DGH instability \cite{Harris1959,Dory1965}.  We show new regimes for the appearance of instabilities and describe energy transfer between ion populations.
For any application, spreads in $k$, inhomogeneity, finite system size, and particle losses will also impose practical constraints and thresholds.
As a secondary distinction, the present root-pole process differs from the lower-hybrid/IBW root-root coupling studied by Kotani and co-workers \cite{Kotani2026}.

\section{ACKNOWLEDGMENTS}
The authors thank Dr. Renaud Gueroult for useful discussions.  H.Q. was supported in part by flexible research funds from the University of Wisconsin–Madison, including funds provided by the Wisconsin Alumni Research Foundation (WARF).   N.J.F. was supported in part by DOE Grant No. DE-SC0016072. 

OpenAI ChatGPT (GPT-6) was used to assist with language editing and verification of analytical derivations and numerical calculations. All scientific results were developed by the authors, who take full responsibility for the content of the manuscript.

\bibliographystyle{apsrev4-2}
\bibliography{references}

\end{document}